\documentclass{epl}

\title{Measurement of the $^3$He mass diffusion coefficient in
superfluid $^4$He over the 0.45-0.95 K temperature range}

\shorttitle{Measurement of the $^3$He diffusion coefficient}

\author{S.K. Lamoreaux\inst{1} \and
G. Archibald\inst{2} \and P.D. Barnes\inst{1} \and W.T. Buttler\inst{1} \and
 D.J. Clark\inst{1} \and  M.D. Cooper\inst{1} \and M. Espy\inst{1} \and 
G.L. Greene\inst{1} \and R. Golub\inst{3},
M.E. Hayden\inst{2} \and C. Lei\inst{2}, L.J. Marek\inst{1} \and
J.-C. Peng\inst{1} \and S. Penttila\inst{1}}
\shortauthor{S.K. Lamoreaux \etal}
\institute{
\inst{1} University of California,
Los Alamos National Laboratory, Physics Division,
Los Alamos, New Mexico 87545, USA\\
\inst{2} Department of Physics, Simon Fraser University, 8888 University
Drive, Burnaby, B.C., Canada V5A 1S6\\
\inst{3} Hahn-Meitner Institut, Glienicker Str. 100,
D-14109 Berlin, Germany
}

\begin{document}

\maketitle

\begin{abstract}
We have measured
the mass diffusion coefficient $D$ of $^3$He
in superfluid $^4$He at temperatures
lower than were previously possible.  The experimental technique
utilizes scintillation light produced when neutrons react
with $^3$He nuclei,
and allows measurement of the $^3$He density integrated along 
the trajectory of a well-defined neutron beam. By measuring
the change in $^3$He density near a heater as a function
of applied heat current,
we are able to infer values for $D$
with 20\% accuracy.
At temperatures below 0.7 K and for concentrations
of order $10^{-4}$, we find $D=\left[2.0 {+2.4 \atop -1.2}\right]
T^{-(6.5\mp 1.2)}$ 
cm$^2$/s, in
agreement with a theoretical approximation.
\end{abstract}

\section{Introduction}

We have developed a new technique for studying the
distribution of $^3$He atoms at low concentrations
in superfluid $^4$He and have
applied it to the 
measurement of the $^3$He mass diffusion coefficient $D$
at temperatures below 1 K. 
Previous measurements of  $D$
have not extended
to temperatures below 1 K \cite{beenaker,ptukha}.  The earlier
results are based on thermal conductivity measurements
and, because $D$ becomes very
large at low temperatures, an impractical amount of heat input
is required to achieve a measurable temperature difference below
1 K.

We are interested in the magnitude of $D$
near and below $0.5$ K, where it is expected to be large \cite{gl}.
Reference \cite{beenaker} indicates that $D=.025\ {\rm cm^2/s}$ at
$T=1.2$ K, and that it increases rapidly with decreasing
temperature; the temperature dependence of
$D$ is predicted to be  $T^{-7}$ for $T<0.6$ K \cite{gl}.

\section{Theory of the experimental technique}

Our method of measuring $D$ is based on a comparison of
the $^3$He distribution in a cell with and without a source of
heat in the liquid.
In the treatment below we consider heat flow and $^3$He concentrations
that are sufficiently small
such that we can assume the liquid is isothermal.
We also ignore the relatively
infrequent $^3$He-$^3$He collisions that
do not influence the steady-state $^3$He atom distribution imposed by the
heat flow.
This system can be analyzed 
in terms of the usual
macroscopic two-fluid model for liquid
$^4$He below the $\lambda$ point \cite{london}; 
these two fluids comprise the superfluid
component that carries no entropy, and the normal component that,
at sufficiently low temperatures,
carries thermal energy as a pure phonon gas.   
When a heat source and a heat sink
are introduced, the heat flux $\bf q$ associated
with the normal component flow is
\begin{equation}\label{normal}
{\bf q}=\rho s T {\bf v_n}
\end{equation}
where $\rho$ is the $^4$He liquid density, $s$ is the entropy per
unit mass, $T$ is the temperature, and $\bf v_n$ is the normal component
velocity.  The $^3$He atoms 
scatter from the phonons and are thus carried along with
the normal fluid flow. Consequently, the $^3$He atoms accumulate
at the heat sink.  The $^3$He atoms also scatter
from the background phonons that exist when the superfluid $^4$He
is at finite temperature; this scattering tends to randomize
the $^3$He distribution.  The net effect is described by the
diffusion equation
\begin{equation}\label{diff}
X{\bf v_n}-D\vec\nabla X=0
\end{equation}
where $X$ represents the spatially varying $^3$He concentration
(see, e.g., \S29, eq. (13) of \cite{london}).
In steady state,
\begin{equation}
{\partial \rho s\over\partial t}=0=-\vec\nabla\cdot\rho s {\bf v_n}.
\end{equation}
We assume $\rho$ and $s$ are spatially uniform
under isothermal conditions in which case
$\vec\nabla\cdot {\bf v_n}=0$.  Furthermore, if
the flow is laminar, 
then $\vec\nabla\times{\bf v_n}=0$, which allows us to write
${\bf v_n}=\vec\nabla \Phi$, where the velocity
potential $\Phi$ satisfies Laplace's
equation, $\nabla^2 \Phi=0$. Substituting this result into eq. (\ref{diff}),
\begin{equation}
X\vec\nabla\Phi-D\vec\nabla X=0\Rightarrow {1\over D}\vec\nabla\Phi=
{1\over X}\vec\nabla X=\vec\nabla \log(X),
\end{equation}
which has the solution
\begin{equation}\label{tr}
X=Ce^{\Phi/D}.
\end{equation}
The coefficient $C$, determined by the requirement
that the total number of
$^3$He atoms  must be constant for a closed system,
depends on the total heat input and $D$.
For the case of a point heat source of power $P$ in a large bath
(i.e., large enough so we can ignore boundary conditions),
\begin{equation}\label{point1}
q(r)={P\over 4\pi r^2}\Rightarrow v_n(r)={q(r)\over \rho s T}\Rightarrow
\Phi(r)=-{P\over \rho s T}{1\over 4\pi r}=-{\alpha\over 4\pi r}
\end{equation}
where $r$ is the distance from the source, and $\alpha =P/(\rho s T)$.
In this case,
\begin{equation}\label{point2}
X(r)=Ce^{-\alpha /4\pi Dr}
\end{equation}
and
$D$ can be determined by measuring 
the spatial extent of the $^3$He-depleted region as a function
of $P$. Alternatively, $D$ can be
inferred if the integrated
$^3$He density along a specific path near the heat source is measured
as a function of $P$.  The latter approach serves as the basis of the
measurements reported here.

\section{Description of the experiment}

Use of a neutron beam to determine tomographically the distribution
of $^3$He atoms subjected to heat currents in
a  superfluid-$^4$He-filled cell is described in Appendix C of \cite{gl}.
For the measurements reported here, we use a variant of this
technique; we determine the differential heat on/off
effect using a fixed and
well-characterized neutron beam.
A scale drawing of our apparatus is shown in Fig. \ref{f.1}.  
The 100 cm$^3$ cell containing superfluid $^4$He is
mounted on the end of a horizontal dilution
refrigerator; the OFHC copper mixing chamber 
of the dilution refrigerator, which serves as a heat sink,
forms
one end of the cell.  The cell body is a cylindrical shell
of aluminum approximately 1 mm thick.  A fused silica window
forms the other end of the cell.

Neutrons are absorbed by $^3$He according to the reaction
\begin{equation}
{\rm ^3He +n\rightarrow p+t +784 keV}
\end{equation}
and the energy released to the proton and triton create XUV
(80 nm) scintillation
light as they are stopped in the superfluid
$^4$He.  The XUV light is wavelength-shifted to 450 nm 
with tetraphenyl butadiene (TPB) evaporated onto the cell walls
and onto an aluminized
mylar sheet that covers most of the copper end of the cell.
The 450 nm light is transmitted through a series of windows. In addition
to the cell window, there are two fused
silica windows at about 1 K, one fused silica window at about 77 K, an
acrylic (blackbody radiation
absorbing) window at 77 K, and a fused silica window at room temperature.
The scintillations are then detected by a Hamamatsu R329-02 photomultiplier,
set to a threshold of $5\pm 0.5$ photoelectrons; the detection efficiency of
$^3$He capture events varies spatially 
within the cell in the 40 to 100\% range, with the variation
depending primarily on the distance from the mylar sheet.

The experiment was performed at
the Los Alamos Neutron Science Center spallation source and used
Flight Path 11 A, which comprises a cold neutron guide that views a
liquid hydrogen moderator.
The neutron beam was formed by collimating
the output of the cold neutron guide to a 2.5 mm FWHM beam with a divergence
less than $1\times 10^{-3}$ rad.  
The pulse structure of the source is not crucial to the 
technique, but was used in our measurements to eliminate the 
background due to the intense 
gamma and energetic charged particle pulse associated with the
proton current pulse incident on the spallation target.
The pulse counters were gated open in a time window of 6 ms following
a delay of 17 ms from the spallation pulse; the distance between
the experiment and moderator was 24 m, so the neutron velocity
range was 1040 to 1410 m/s.  Approximately 3\% of the incident
neutrons are scattered by the superfluid $^4$He, and at our
level of accuracy, the effects due to scattered incident
neutrons can be neglected.

The initial 
$^3$He concentration $X_0$ was varied during the course of the experiment
such that between 5 and 20\% of the incident neutrons were absorbed,
corresponding to $7\times 10^{-5}\leq X_0\leq 3\times 10^{-4}$.
The gated count
rate was typically 450/s ($X_0=10^{-4}$) with a background of 12/s.
The background was determined by operating the experiment with
the cold neutron component of the beam blocked by thin Cd
and $^6$Li doped plastic sheets, and was independent of heat
input and relatively insensitive to the $^3$He concentration.
This background was largely due to long-lived molecular
optical excitations in the
superfluid bath \cite{klaus} that were left over from the spallation
radiation flash.  In addition, the neutron beam intensity transmitted
through the cell was measured with a $^6$Li-doped glass scintillator
detector, gated identically with the helium scintillation signal.
The proton pulse current provided by the facility was used to
obtain a relative normalization of the cold neutron intensity.

In order to perform tomography, the cryostat containing
the dilution refrigerator and cell was placed on a motor-controlled
table that allowed horizontal and vertical motion perpendicular
to the neutron beam.  Linear resistive displacement transducers
linked to the table motion
allowed positioning to 0.5 mm absolute accuracy, and to better than 0.25 mm 
reproducibility.  The heat source
was a carbon composition  resistor (50 $\Omega$ at room temperature)
located near the bottom of the cell and supported by an insulating
alumina column.  A calibrated ruthenium oxide temperature-sensing resistor 
supported by its wire
leads was located near the mixing chamber end of the cell as
shown in Fig. \ref{f.1} This resistor was
used in conjunction with a heater located in the
mixing chamber to provide active relative
temperature stabilization to within 20 mK under varying
heat input conditions. 
The uncertainty in the absolute temperature  was
about 0.03 K and depended on the operating temperature.  The
maximum power delivered to the cell was in the 7 to 15 mW 
range and was limited by the temperature-dependent cooling
power of the refrigerator.  

The helium scintillation signal provides a relative measure of
the volume-integrated $^3$He density
along the neutron beam.  In order to interpret the effects of
the applied heat flux it is necessary to
determine the velocity potential $\Phi(x,y,z)$ within the 
cell.  This was done by
numerically solving $\nabla^2\Phi=0$
on a $100\times 100\times 100$ grid using
standard relaxation techniques, with the boundary condition
that the perpendicular component of $\bf v_n$ was zero at all
boundaries, except at the mixing chamber,
where the condition $\Phi=0$ was set. The perpendicular
component of $\bf v_n$
at the resistor surface was determined by the total heat input
and assumed to be uniform.  Finally,  the normalization was
determined by integrating $e^{\Phi}$ over the grid,
with $D$ left as a parameter (scale factor) to be
determined experimentally. The
integrated $^3$He density was determined by evaluating
the integral of $X$ over the neutron,
as beam shown in Fig. \ref{f.1}.

We verified experimentally
that eq. (\ref{tr})  does indeed describe the
distribution of $^3$He in the cell subject to heat flow
(to be published).  For 
determining $D$, the beam was directed through
the cell
in the region near the heater; the beam location used
for the measurements is indicated in Fig. \ref{f.1}.  
Results of
the numerical calculation of the integrated $^3$He 
density along the neutron path
shown in Fig. \ref{f.1}  are given in Fig. \ref{f.2} as a function
of heat power; the value of $D$
at a given temperature enters as an unknown scale factor on the
horizontal axis of Fig. \ref{f.2}.  By measuring the
relative rate of change of scintillation rate $R$ (which
is proportional to the relative change in integrated $^3$He density)
near zero power,
$D$ can be determined from
\begin{equation}\label{slope}
D={-0.078\ {\rm cm^{-1}}\over dR/d\alpha},
\end{equation}
where the numerical factor is  the maximum
slope (i.e., at zero heat power) in Fig. \ref{f.2}.
Because we expect that $D$ scales as $T^{-7}$ at low temperatures
(see below),
we can cast the data in terms of a universal function if we
plot the experimentally-observed 
relative change in scintillation rate as a function
of $x=\alpha T^7$. Then eq. (\ref{slope}), with
$\alpha$ replaced by $x$, yields $D_T$
directly, where $D=D_T T^{-7}$.

\section{Experimental Results}

The data set is shown in Fig. \ref{f.3}.  The
relative scintillation rate  $R(x)/R(0)$ 
as a function of heater power (parameterized by $x$)
was determined by subtracting
the background rate from the observed scintillation rate, 
and dividing by the background-subtracted
scintillation rate at zero power.  The vertical errors
were determined by counting statistics, while the
horizontal errors were determined by
the estimated uncertainties in $T$ (and therefore $s$). 
The low
temperature points tend to fall on the same curve; the
small values of $T^7$ at the lower temperatures, along with the
limited cooling power of the dilution refrigerator, restricted these
data to $x\leq 2.2$ as shown in Fig. \ref{f.3}. 
This is sufficiently small so that the
variation in relative scintillation rate can be assumed to be 
linear over the range $0\leq x\leq 2.2$ (see Fig. \ref{f.2}).  The
low temperature points (e.g., the data below
0.7 K) are re-plotted in Fig. \ref{f.4} using
expanded axes.  The points fall on a single
line with reduced $\chi^2=0.78$.  From the slope of this line, along with
eq. (\ref{slope}),  we arrive at
\begin{equation}\label{result}
D=D_T T^{-7}=(1.6\pm 0.2)T^{-7}.
\end{equation}
As expected, there was no
observed effect due to concentration (e.g., the low
temperature data are consistent as indicated by $\chi^2$),
so we
present no analysis on the effect of varying  $X_0$.
Also,  from eq. (\ref{normal}),
 $v_n<10$ cm/s for all the data; this is sufficiently
small so that turbulence effects are unlikely to
be important.

Finally, $D$ was determined at each temperature independently
as shown in Fig. \ref{f.5}, along with results from previous
work.  It can be seen that our results extrapolate
to the Beenakker et al.  \cite{beenaker} results, but both experiments
are inconsistent with the Ptukha  \cite{ptukha} results.
One can further
analyze the lowest-temperature
data in a two-dimensional fit using $D=D_TT^{-N}$
yielding
\begin{equation}
D=\left[2.0 {+2.4 \atop -1.2}\right]
T^{-(6.5\mp 1.2)}\ \ {\rm cm^2/s}
\end{equation}
and indicates the level at which the theory presented
below is constrained by the
experiment.

\section{Microscopic theory of $^3$He impurity diffusion}

The expectation that $D$ will vary as $T^{-7}$ when collisions with
phonons provide the dominant $^3$He scattering mechanism ($T<0.6$ K) 
was briefly discussed in
Ref.  \cite{gl} (Sec. 7.2); here we add further details to that
argument. From Fig. 2 of \cite{husson} we infer that the $^3$He-phonon
collision frequency
$1/\tau_{ph}$ varies as $T^4$. The data at $T=0.5$ K and
concentration $X=1\times 10^{-3}$ in this figure imply
\begin{equation}
1/\tau_{ph}=4.8\times 10^{10}\ X\ T^4\ {\rm s^{-1}}.
\end{equation}
Because the number of collisions per unit volume per unit time
must be the same for phonons and $^3$He atoms, we have
\begin{equation}
{n_{ph}\over \tau_{ph}}=
{n_3\over \tau_3}\Rightarrow {1\over \tau_3}={n_{ph}\over n_3}
{1\over \tau_{ph}}
\end{equation}
where $n_3=2\times 10^{22}\ X\ {\rm cm^{-3}}$.
The phonon density is  approximately
(e.g., \cite{london}, \S16, eq. (28))
\begin{equation}
n_{ph}=8\pi\zeta(3)(kT/hc)^3=2\times 10^{19} T^3\ {\rm cm^{-3}}
\end{equation}
where $\zeta(3)=1.202$,
$c=2.4\times 10^{4}\ {\rm cm/s}$ is the phonon phase velocity,
$k$ is Boltzmann's constant, and $h$ is Planck's constant; therefore
\begin{equation}
{1\over \tau_3}=4.8\times 10^7 T^7 {\rm s^{-1}}.
\end{equation}
The collisions measured in Ref. \cite{husson} are those effective
in phonon transport, i.e., those collisions that change the phonon
momentum by a significant amount.  For elastic collisions, which
dominate here, the momentum transfer for a phonon of momentum
$q$ scattered through an angle $\theta_q$ (where $\Delta q=2q\sin\theta_q/2$),
is equal to the momentum transfer to a $^3$He atom with momentum
$p_3$, scattered through an angle $\theta_{3}$ 
(where $\Delta p_3=2p_3\sin\theta_{3}/2$),
in a given collision.  For each collision,
\begin{equation}
p_3^2(1-\cos\theta_3)=q^2(1-\cos\theta_q)\Rightarrow \langle 
1-\cos\theta_3\rangle
\approx \langle 1-\cos\theta_q\rangle\langle q^2\rangle/\langle p_3^2\rangle;
\end{equation}
where $\langle...\rangle$ indicate a thermal
average.  From elementary considerations, 
the average squared momenta are
\begin{equation}
\langle q^2\rangle\approx
{\langle E^2_{ph}\rangle \over c^2}
\approx {10.4 (kT)^2\over c^2}, 
\ \ \ \ \ \ \langle p_3^2\rangle={3kTm_3^*}
\end{equation}
where
$m_3^*=2.2m_3$ is the effective $^3$He mass 
in superfluid $^4$He ($m_3$ is the $^3$He atomic mass).  Numerically, 
$\langle q^2\rangle/\langle p_3^2\rangle=7.5\times 10^{-2} T$
so for $T<1$ K, $p_3^2>>q^2$ and $\langle \cos\theta_q\rangle\approx 0$.
From kinetic theory,  
the diffusion coefficient is proportional to the product
of $1/\langle 1-\cos\theta_3\rangle$ and
$\langle v_3 \lambda_3\rangle=\tau_3\langle v_3^2\rangle$,
where $v_3$ is the $^3$He velocity and
$\lambda_3=v_3\tau_3$ is the distance a $^3$He atom
travels between collisions: 
\begin{equation}
D={1\over 3} \langle\lambda_3 v_3\rangle/\langle1-\cos\theta_q\rangle=
{1\over 3} \langle v_3^2\rangle\tau_3/
\langle 1-\cos\theta_q\rangle.
\end{equation}
We find that
\begin{equation}\label{theory}
D=1.2\ T^{-7} {\rm cm^2/s} = D_T T^{-7}.
\end{equation}
The factor $D_T$ is likely correct to within an order
of magnitude and is applicable to temperatures
near and below 0.6 K, e.g., where the thermodynamic
properties of superfluid $^4$He are dominated by
phonons. More importantly, however,
this result indicates the functional form of $D$ at low
temperature.
The value of $D_T$ from eq. (\ref{theory}) is in
reasonable agreement with our experimental result, 
eq. (\ref{result}). 
A useful figure-of-merit for the effectiveness of
a given heat input power for
transporting $^3$He atoms by
phonon-dominated entrainment is, by use of 
eqs. (\ref{point1}) and (\ref{point2}),
$sT/D \propto T^{5}$ 
($s\propto T^{-3}$ for $T<0.7$ K).
The deviations from the  low temperature
$T^{-7}$ behavior of $D$ that are evident for $T>0.65$ K
in Fig. \ref{f.5} are likely due to the influence of 
of rotons at higher temperatures.

\section{Conclusion}

We have employed a novel differential
absorption 
technique to determine the mass diffusion coefficient of 
$^3$He in superfluid $^4$He as a function of temperature.
The measurements extend to much lower temperature than
in any previous work. We have verified that the diffusion
coefficient is closely proportional to $T^{-7}$ at temperatures
below 0.7 K.

This work was supported by Los Alamos National Laboratory
LDRD under Project Numbers 97041, 98039, and 2001526DR.

\begin{figure}
\onefigure[scale=0.6]{fig1.eps}
\caption{Scale drawing of the test cell. There are
indium seals at the connections between the aluminum cell
cylinder, demountable window, and dilution refrigerator.
The cell is filled by condensing the desired
mixture of helium isotopes through the fill tube.
The 50 $\Omega$ resistor serves as the heater.}
\label{f.1}
\end{figure}
\begin{figure}
\twofigures[scale=0.4]{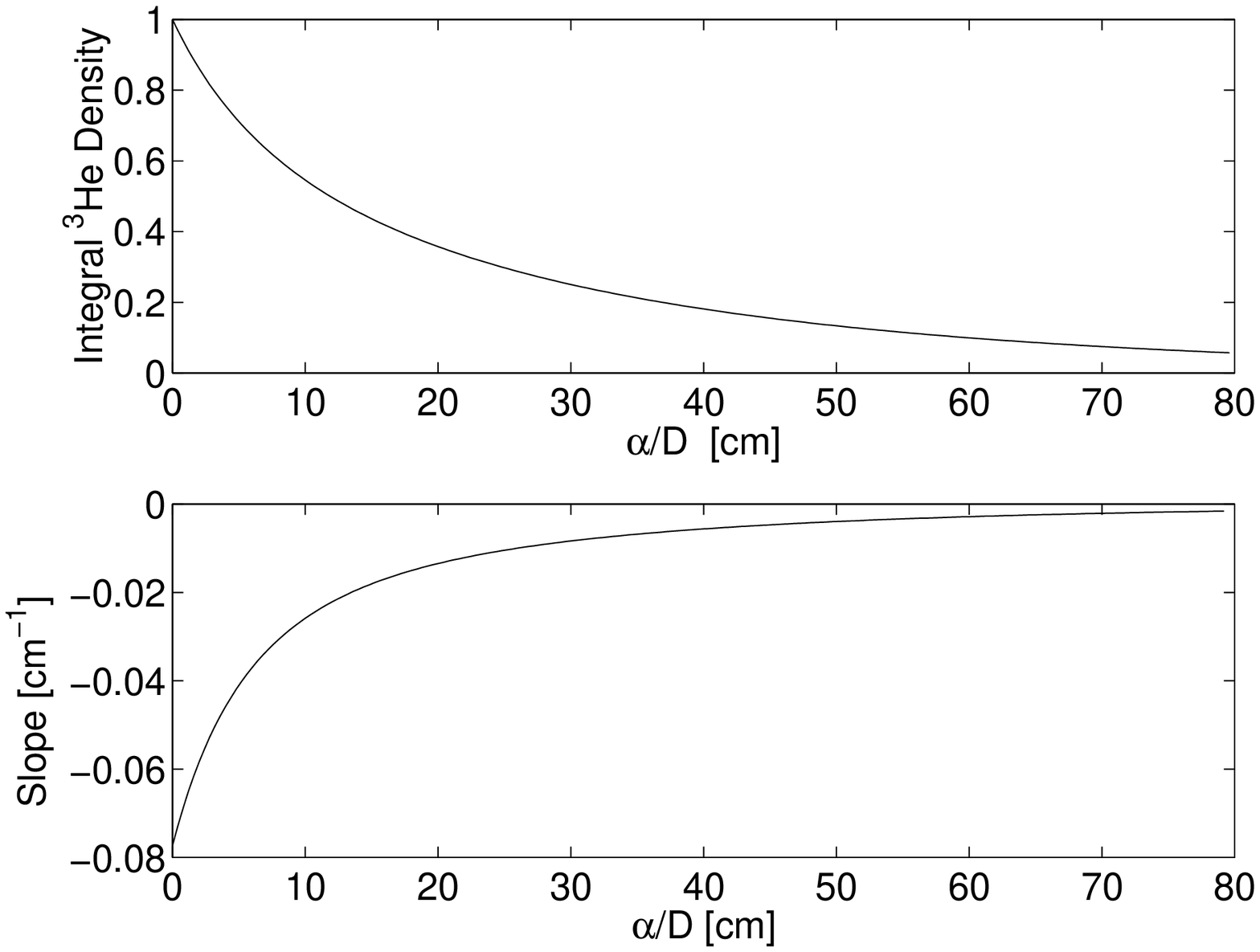}{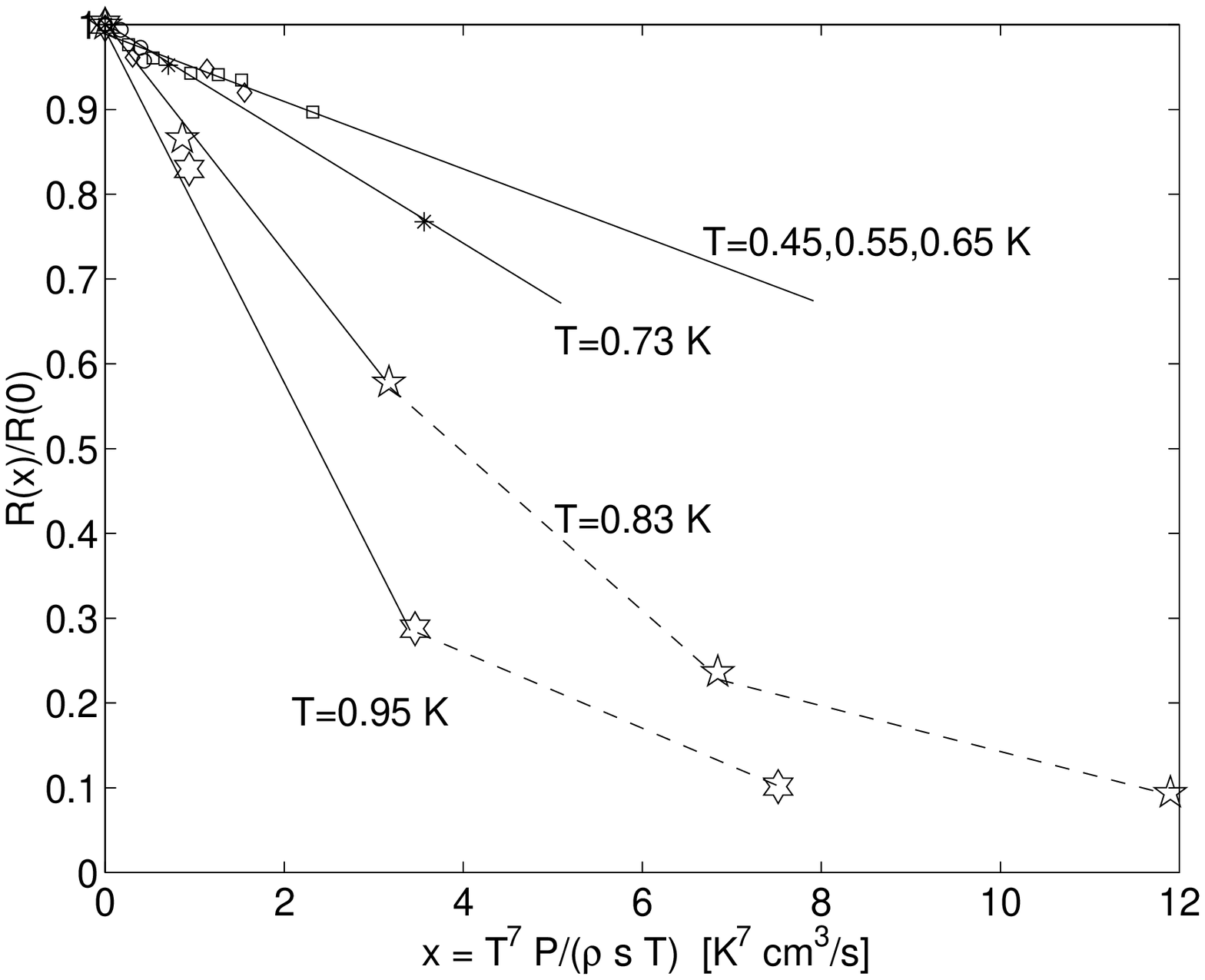}
\caption{Numerical results for 
the relative change in the integrated $^3$He density
integrated along the the
neutron beam path
as a function of input power $P$, parameterized as $\alpha/D$; 
the lower plot shows the numerical derivative of the
upper plot. These results
are specific to the geometry of our experiment.}
\label{f.2}
\caption{Experimentally-determined relative change
in scintillation rate data plotted as a function of
$x=\alpha T^7$.
The data points for $R(x)/R(0)> 0.6$ represent a total count of
about 30,000 scintillation
pulses, from which a background of 1000 has been
subtracted; the background is independent of
$^3$He density and heater power. The errors in temperature and relative
counting statistics are of order the size of the symbols
used to plot the data. Solid lines indicate the slope at low
heater power, while the dashed lines indicate the trends in the data at
high power.}
\label{f.3}
\end{figure}
\begin{figure}
\twofigures[scale=0.4]{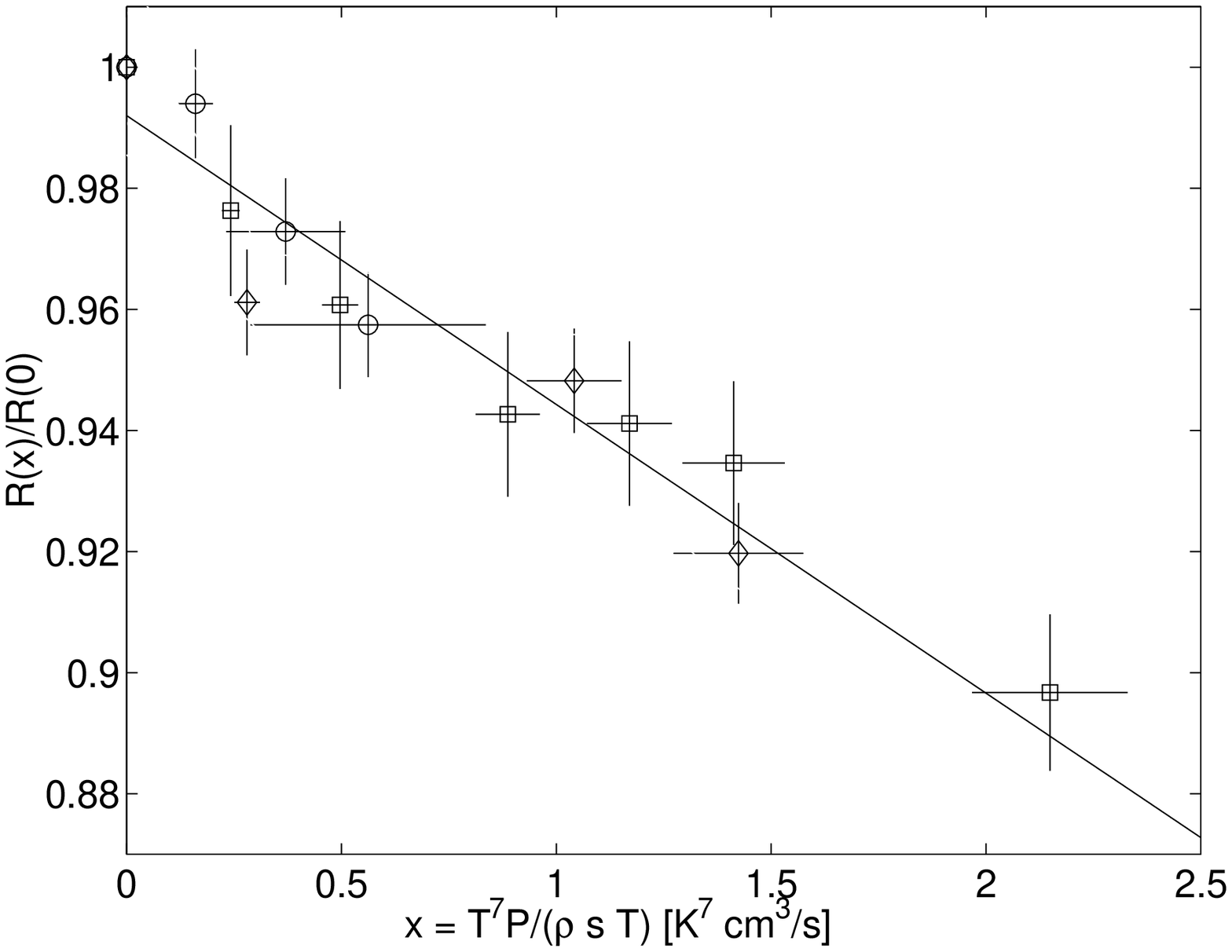}{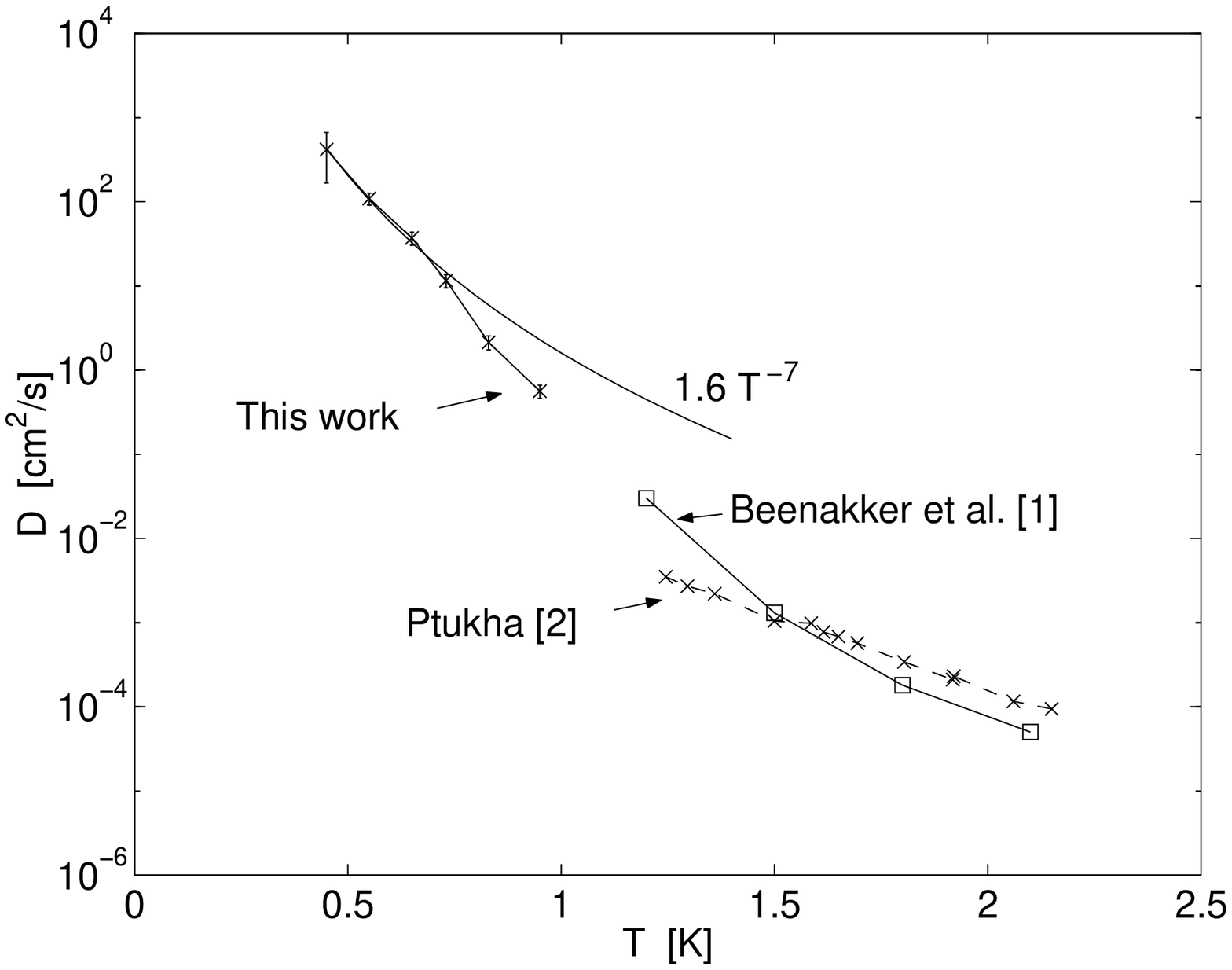}
\caption{An expanded view of the low temperature
data from Fig. \ref{f.3}., which fall on a
single line the the low power regime
(circles, 0.45 K; diamonds,
0.55 K; squares, 0.65 K).
This behavior indicates that $D$ scales as $T^{-7}$.  A least-squares
fit to a 
line $y=mx+b$ gives $b=0.992\pm .004$, $m=(-4.77\pm 0.45)\times
10^{-2} {\rm s/cm^3K^7}$, with a reduced $\chi^2=0.78$.  }
\label{f.4}
\caption{A comparison of our results for the mass diffusion
coefficient $D$ of $^3$He in superfluid $^4$He below 1 K with those
of previous workers at higher temperature. }
\label{f.5}
\end{figure}

\end{document}